\documentclass[aps,prb,showpacs,twocolumn,reprint,superscriptaddress,showkeys,longbibliography]{revtex4-1}
\usepackage{amsmath}
\usepackage{amssymb}
\usepackage{graphicx}
\usepackage{color}
\usepackage{multirow}
\usepackage{dcolumn}
\usepackage{textcomp}
\usepackage{longtable}
\usepackage{makecell}
\begin{document}
\title{Prediction of giant and ideal Rashba-type splitting in ordered alloy monolayers grown on a polar surface}
\author {Mingxing Chen}
\email{mxchen@hunnu.edu.cn}
\affiliation{
School of Physics and Electronics, Hunan Normal University, 
Key Laboratory for Matter Microstructure and Function of Hunan Province,
Key Laboratory of Low-Dimensional Quantum Structures and Quantum Control of Ministry of Education,
Changsha, Hunan 410081, China}
\author {Feng Liu}
\affiliation{
Department of Materials Science and Engineering, University of Utah, Salt Lake City, Utah 84112, USA}
\date{\today}

\begin{abstract}
A large and ideal Rashba-type spin-orbit splitting is desired for the applications of materials 
in spintronic devices and the detection of Majorana Fermions in solids.
Here, we propose an approach to achieve giant and ideal spin-orbit splittings 
through a combination of ordered surface alloying and interface engineering, 
that is, growing alloy monolayers on an insulating polar surface.
We illustrate this unique strategy by means of first-principles calculations of buckled hexagonal monolayers of 
SbBi and PbBi supported on Al$_2$O$_3$(0001).
Both systems display ideal Rashba-type states with giant SO splittings,
characterized with energy offsets over 600 meV and momentum offsets over 0.3 $\AA^{-1}$, respectively. 
Our study thus points to an effective way of tuning spin-orbit splitting in low-dimensional materials to draw immediate experimental interest.
\end{abstract}

\keywords{Rashba; interface; spin-orbit coupling}

\maketitle
\section*{INTRODUCTION}
The Rashba effect is referred to as the spin-orbit (SO) splitting at surfaces/interfaces due to the broken inversion symmetry\cite{Rashba,LAO_STO} , 
which has led to many exotic quantum phenomena and novel applications,
ranging from spin Hall effect, Majorana fermions in solids and the spin field-effect transistor\cite{PRL_SHE,Qi2011,Datta-Das}.
The effective Rashba Hamiltonian for an electron with momentum $\mathbf{k}$ and spin $\mathbf{\sigma}$ can be written as 
\begin{equation*}
 \hat{H}_{R} = \lambda \mathbf{\sigma} \cdot (\mathbf{E_z}\times\mathbf{k}),
\label{eq1}
\end{equation*}
where $\lambda$ is the strength of SO coupling (SOC) and
$\mathbf{E_z}$ the electric field perpendicular to the surface/interface 
is created by a perpendicular potential gradient related to the structural asymmetry.
The SO splitting is defined as $\alpha_R = 2E_R/k_R$,
where $E_R$ and $k_R$ are the Rashba energy and momentum offset, respectively.

In the ongoing quest for exploration of large Rashba-type SO splittings in materials, 
 enhancing the strength of $\lambda$ by introducing heavy elements has been extensively used 
\cite{Bi-Ag111,Pb-Ag111,BiPb_2008,BiPb_2009,BiSb_BiPb_2011,Bi-Cu111_2009}.
In particular, the ordered ($\sqrt{3}\times\sqrt{3}$) superstructure of Bi/Ag(111) 
in which one surface Ag is replaced by Bi displays
a giant Rashba energy offset ($E_R$) of 200 meV and momentum offset ($k_R$) of 0.13 $\AA^{-1}$.\cite{Bi-Ag111}
In addition, semiconducting substrates are used to avoid the mixing of the Rashba states and the spin-degenerate substrate states,
in order to create so-called ideal Rashba states
\cite{Tl-Si111-2009,Bi-Si111_2010,Tl-Si111_2014,Tl-Ge111_2008,Pb-Ge111_2010,BiNa_TlPb_2014,TlPb_2015,BiTl_2016,BiSb_2017,Sn2Bi_2018,Au-InSe_2018}. 
On the other hand, interfacial dipole field can be used to enhance $\mathbf{E_z}$ and hence the SO splitting.
Polar semiconductors are effective substrates to serve this purpose
\cite{Au-InSe_2018,Bi_BaTiO3_2010,Bi_BaTiO3_2017,BTO_BRO_2015,LAO_STO_2010,LAO_STO_2019}.
In fact, the surfaces of polar semiconductors bismuth tellurohalides BiTeX (X = Cl, Br, and I) have been found to exhibit giant SO splittings
\cite{BiTeCl,BiTeX}.
Moreover, the Rashba SO splitting can be controlled by the electric polarization in ferroelectric materials and substrates
\cite{Bi_BaTiO3_2010,Bi_BaTiO3_2017,BTO_BRO_2015,GeTe,GeTe_2017,PbTe}. 
Despite these achievements, natural materials exhibiting both giant and ideal Rashba states are rare.
Therefore, artifical interfaces by \textit{a priori} theoretical design is highly desirable to fill this outstanding gap.

In this work, we demonstrate the design principle to creat giant and ideal interfacial Rashba states
by combining ordered surface alloying and growth on a polar insulator/semiconductor surface.
Using density-functional theory (DFT) calculations, we show unprecedented large Rashba energy offsets over 600 meV 
and momentum offsets over 0.3 $\AA^{-1}$ for both SbBi/Al$_2$O$_3$(0001) and PbBi/Al$_2$O$_3$(0001), 
which are roughly three times of those in Bi/Ag(111). 
Also, such Rashba states are ideally situated inside the band gap of Al$_2$O$_3$(0001). 

\section*{RESULTS}

\begin{figure}
  \includegraphics[width=0.45\textwidth]{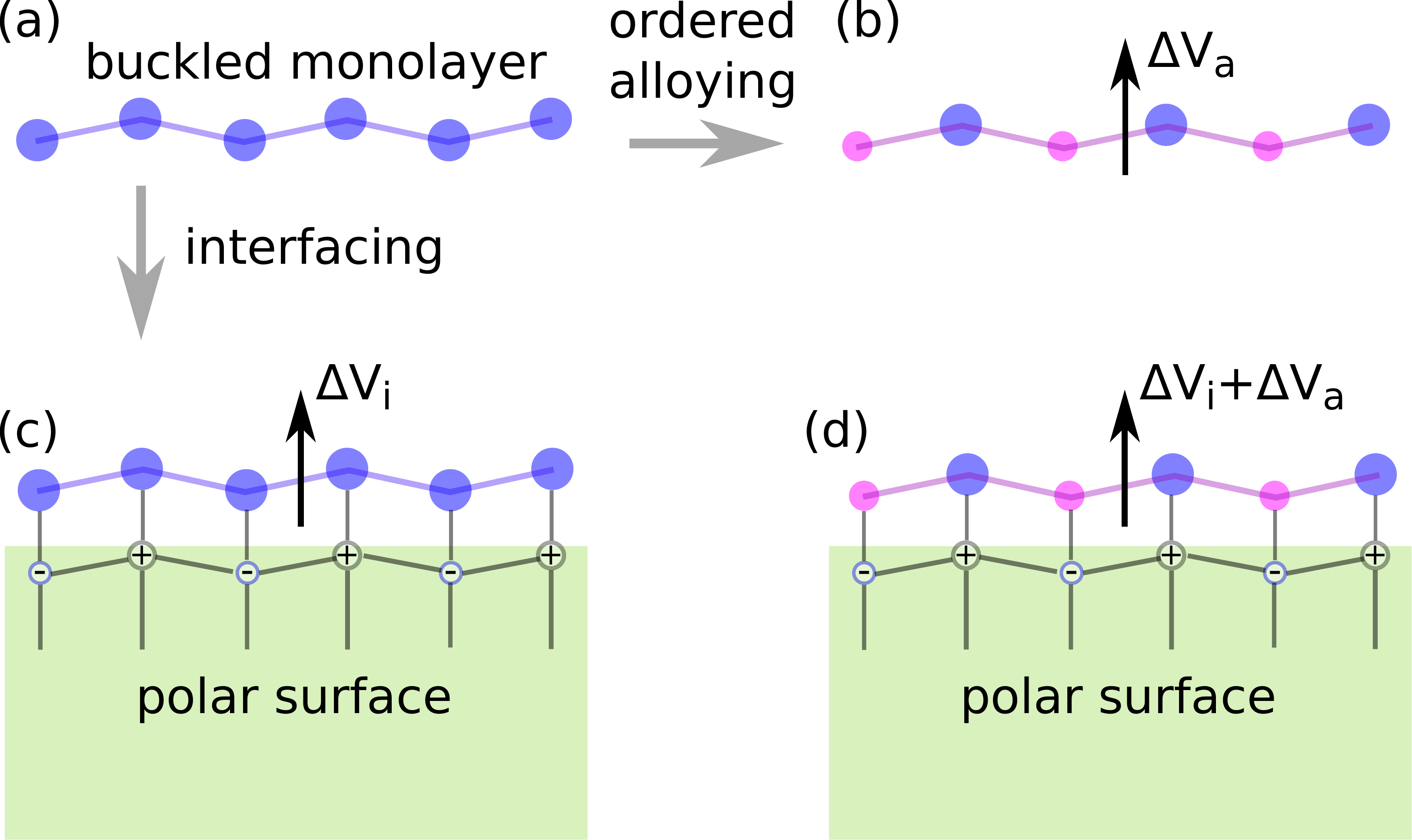}
  \caption{Sketch of manipulating the potential gradient through a buckled monolayer by ordered alloying and interfacing.
  (a) The geometry of a free-standing buckled monolayer.
  (b) Perpendicular potential gradient ($\Delta V_a$) induced by ordered alloying.
  (c) Interface-induced potential gradient ($\Delta V_i$) through the monolayer when placing it onto a polar surface, 
  e.g., growing it on a substrate.
  A large $\Delta V_i$ can be favored when they form a special interface structure such that 
  the two atoms in the monolayer bind to different types of ions in a polar surface.
  (d) The combined effect of the ordered alloying and interfacing.
  The arrows indicate the perpendicular potential gradients.
  Different types of atoms are in different colors.
  The symbol +/- represents cations/anions in the polar surface.
  }
 \label{fig1}
\end{figure}

We begin by illustrating the general idea as shown in Fig. 1.
Our structural model makes use of the geometric and electronic properties of both the buckled overlayer and substrate.
The buckled honeycomb structure exists in a number of elemental layered materials, 
such as silicene, germanene, stanene, and Bi(111) monolayers
\cite{silicene,stanene,Bi111,Wada_2011,Liu_Bi111_2011},
which consists of two trigonal sublattices sitting at different heights.
Now imagine these two sublattices are made of different types of atoms to break inversion symmetry. 
Then a Rashba SO splitting will arise in such a buckled honeycomb alloy monolayer. 
Apparently, to enhance SOC, the optimal choice to form the monolayer are heavy atoms, such as Bi, Pb and Sb. 
This constitutes our first idea of surface alloying effect. 
Next, let us imagine to grow this alloy monolayer on a polar insulator/semiconductor substrate.
The polar surface induces an additional perpendicular potential gradient ($\Delta V_i$) through the alloy monolayer.
If $\Delta V_i$ is along the same direction as the alloying induced potential gradient $\Delta V_a$, 
then it will further enhance $\Delta V_a$ (see Fig.~\ref{fig1}d).
Consequently, the combined effects of surface alloying and polar surface conspire leading to a giant SO in the monolayer.
By properly choosing the monolayer-substrate materials combinations, 
one can further tune the relative positions of monolayer SO states relative to that of substrate band gap to achieve ideal Rashba-type states. 

To validate our idea we performed DFT calculations 
for buckled monolayers of Bi (Bi-1L), Pb (Pb-1L), Sb and their ordered alloys on Al$_2$O$_3$(0001).
Al$_2$O$_3$(0001) is chosen for several reasons.
First, it has been extensively used as a substrate for the growth of various materials.
Second, there are two different types of atoms in the surface, i.e., Al and O, which behave chemically different.
In addition, the surface Al atom is slightly higher ($\sim$ 0.15 \AA) than O atom, which is beneficial for growing the buckled monolayers.
Therefore, one may expect an enhanced SO splitting when two different atoms in the overlayer bind to Al and O, respectively.
Moreover, it has a large band gap, which is favorable for forming ideal Rashba states.

\begin{figure*}
  \includegraphics[width=0.95\textwidth]{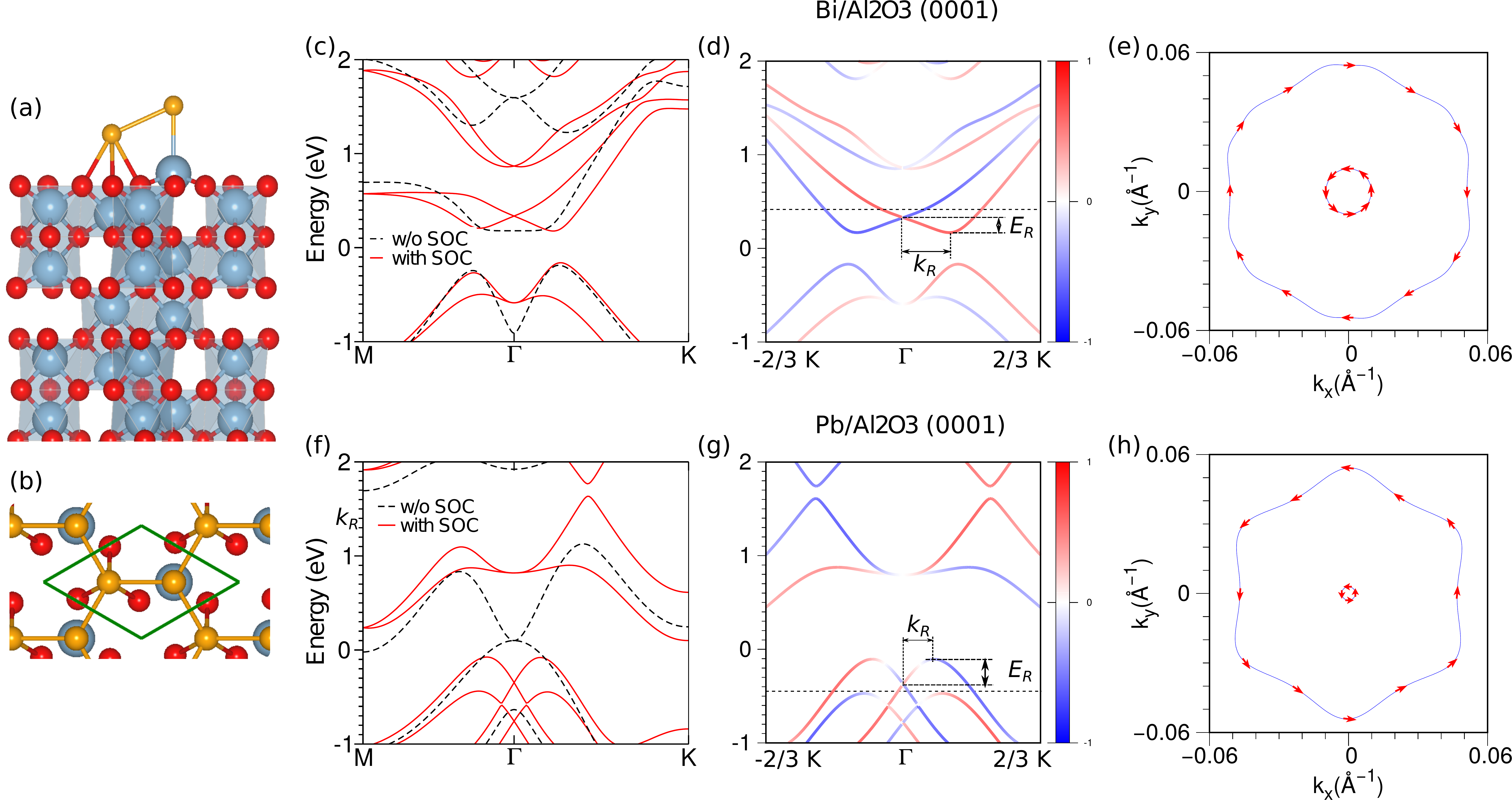}
  \caption{SO splittings in Bi/Al$_2$O$_3$(0001) and Pb/Al$_2$O$_3$(0001).
  (a), (b) Side and top views of the the monolayer supported on Al$_2$O$_3$(0001), 
  The Al and O atoms are denoted by silver and red balls, respectively.
  The green box represents the primitive cell of the interface structure.
  (c), (f) The band structures of Bi/Al$_2$O$_3$(0001) and Pb/Al$_2$O$_3$(0001), respectively.
  The red solid lines denote the bands derived from SOC calculations, 
  whereas the black dashed lines represent the bands from nonrelativistic calculations.
  (d) Spin projections onto the direction vector of $\textbf{K} \times \textbf{e}_z$ for bands along -K-$\Gamma$-K,
  where $\textbf{e}_z = (0, 0, 1)$, for Bi/Al$_2$O$_3$(0001).
  Blue and red represent positive and negative spin polarizations, respectively.
  (e) The Fermi surfaces for the energy marked in (b) by the dashed line.
  (g),(h) Corresponding plots for Pb/Al$_2$O$_3$(0001).
  The red arrows in (e) and (h) denote the spin polarizations.
  $E_R$ and $k_R$ are marked for the bands discussed in the text.
  The Fermi level is set to zero. 
  }
 \label{fig2}
\end{figure*}

We have systematically evaluated the structures and energetics of our systems.
The low-energy structures were derived from our previous calculations of stanene/Al$_2$O$_3$(0001)
using the CALYPSO structure prediction method\cite{CALYPSO},
which produced the same structure for stanene/Al$_2$O$_3$(0001) as reported by Ref.\onlinecite{Sn_Al2O3}.
For the homonuclear monolayers, the structure is shown in Fig.~\ref{fig2}.
The overlayers preserve the buckled honeycomb structure upon the geometric relaxation and 
have a strong binding with the substrate.
The structural properties and energetics are given in Fig. S1.

We first discuss the interfacing effect on the SO splitting in homonuclear monolayers, i.e., Bi/Al$_2$O$_3$(0001) and Pb/Al$_2$O$_3$(0001).
Figure \ref{fig2} shows the band structures for Bi/Al$_2$O$_3$(0001) and Pb/Al$_2$O$_3$(0001) with and without SOC, respectively,
which reveals that both are semiconductors with a gap of about 0.30 eV when SOC is included.
The most prominent feature is the large SO splittings in the overlayer due to the presence of substrate
compared to those of the free-standing monolayers\cite{Pb_2016,Liu_Bi111_2011,Wada_2011}.
For Bi/Al$_2$O$_3$(0001), the conduction band shows a Rashba-like splitting.
While for Pb/Al$_2$O$_3$(0001), the Rashba-like splitting appears in the valance band.
For Pb/Al$_2$O$_3$(0001), there are two series of Rashba-type bands mixed near $\Gamma$.
They become distinct by increasing the layer distance (Fig. S2).
Orbital-projections reveal that the bands near the gap are basically contributed by the $p$-orbitals of Bi (Pb) (Fig. S3).
Thus, the Rashba states are ideal in Bi/Al$_2$O$_3$(0001) and Pb/Al$_2$O$_3$(0001).

We have checked the spin texture of the Rashba states by plotting
spin projections onto the direction vector of $\textbf{K} \times \textbf{e}_z$ for the bands of the two systems, 
where $\textbf{K} = (1/3, 1/3, 0)$ and $\textbf{e}_z = (0, 0, 1)$ is the unit vector normal to the surface. 
The in-plane component perpendicular (parallel/antiparallel) to the vector is denoted by $S_{\perp}$ ($S_{\parallel}$), 
while the out-of-plane component is denoted by $S_z$. 
Figs.~\ref{fig2} (b), (e) show the bands weighted by $S_{\parallel}$ along -K-$\Gamma$-K and  
Figs.~\ref{fig2} (c), (f) show the Fermi surface and spin texture for the energy marked in (b) and (c), respectively.
For the conduction band of Bi/Al$_2$O$_3$(0001), $S_z$ and $S_{\perp}$ are negligible, confirming the Rashba nature of this band,
although the outer branch undergoes a slight warping compared with the inner one.
For the valance band of Pb/Al$_2$O$_3$(0001), 
the warping becomes more prominent and $S_z$ is relatively more appreciable than those for Bi/Al$_2$O$_3$(0001).
The warping indicates that the SO splitting is anisotropic, 
which has the largest value along $\Gamma$-K.
We marked the Rashba energy offset $E_R$ and momentum offset $k_R$ in Figs.~\ref{fig2} (b) and (e).
For Bi/Al$_2$O$_3$(0001), $E_R$ is about 160 meV, comparable to that for Bi/Ag(111) (~200 meV).
Such a SO splitting is one order of magnitude larger than that for the pure Bi(111) (~10 meV) \cite{Bi111_2004}.
While $k_R$ is about 0.2 $\AA^{-1}$, which is 50\% larger than that for Bi/Ag(111).
For Pb/Al$_2$O$_3$(0001), $E_R$ is about 260 meV, which about 30\% larger than that for Bi/Ag(111).
The enhancement in $E_R$ leads to a larger $\alpha_R$ for Pb/Al$_2$O$_3$(0001) compared to Bi/Ag(111),
since $k_R$ is basically the same for both.

\begin{figure}
  \includegraphics[width=0.45\textwidth]{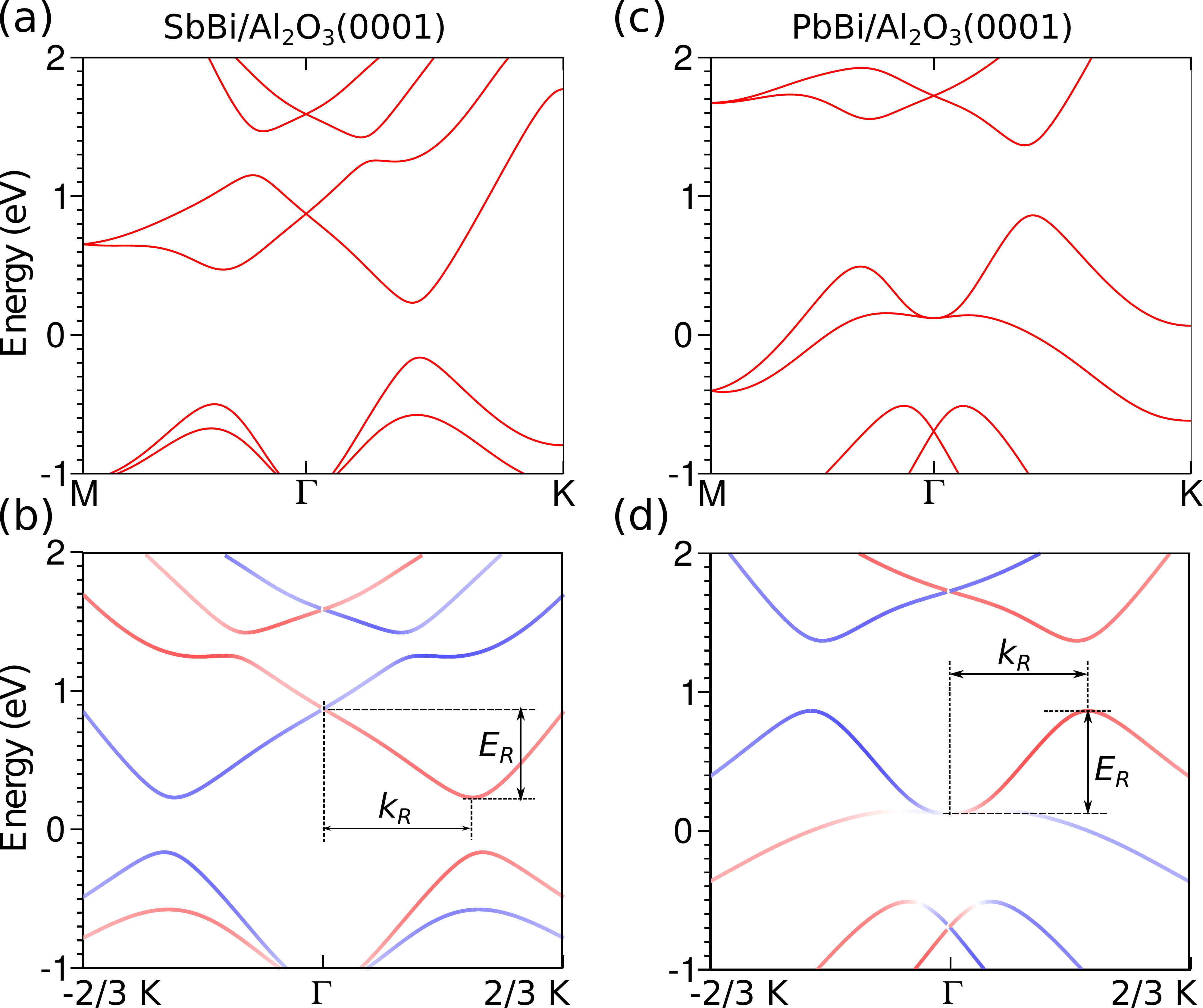}
  \caption{SO splittings in ordered alloy monolayers on Al$_2$O$_3$(0001).
  (a),(c) Band structures of SbBi/Al$_2$O$_3$(0001) and PbBi/Al$_2$O$_3$(0001), respectively.
  (b), (d) Corresponding spin projections onto the direction vector of $\textbf{K} \times \textbf{e}_z$ for bands along -K-$\Gamma$-K.  
  The Fermi level is set to zero. 
  }
 \label{fig3}
\end{figure}
The large SO splitting can be further enhanced by appropriate ordered alloying. 
We substitute one Bi in Bi-1L by an Sb atom, which allows us to obtain a semiconducting monolayer,
while replacing one Bi by Pb leads to split states crossing the Fermi level.
The band structures for the lowest-energy structure for SbBi/Al$_2$O$_3$(0001) and PbBi/Al$_2$O$_3$(0001) are shown in Fig.~\ref{fig3}.
Note that the splitting is now strongly anisotropic compared to those for the homonuclear monolayers.
For SbBi/Al$_2$O$_3$(0001), the bands are pretty much similar to those for Bi/Al$_2$O$_3$(0001).
However, $E_R$ is increased to about 640 meV and $k_R$ is about 0.36 $\AA^{-1}$ for the conduction band along $\Gamma$-K.
Likewise, we obtain a giant SO splitting for PbBi/Al$_2$O$_3$(0001), for which
the energy offset is as large as 740 meV and $k_R$ is about 0.34 $\AA^{-1}$ for the bands crossing the Fermi level.

The large SO splitting is further confirmed by hybrid density functional calculations (see Fig. S4).
We further demonstrate that the large SO splittings are maintained even the chemical composition is not ideal.
To show this effect, we have performed calculations of Sb$_{1.1}$Bi$_{0.9}$/Al$_2$O$_3$(0001) 
and Sb$_{0.9}$Bi$_{1.1}$/Al$_2$O$_3$(0001) using the virtual crystal approximation
to mimic a random alloy. 
Our calculations reveal that 10\% deviation from the ideal chemical composition has only minor effect on the SO splitting (see Fig. S5).
Moreover, to see the effect of the buckling height on the SO splitting, 
we carried out two additional calculations for SbBi/Al$_2$O$_3$(0001) with two different buckling heights. 
The structures were obtained by artificially adjusting the z value of Sb by $\pm$0.1 \AA, which are about 8\% changes in the buckling height.
As a result, there are about 10\% changes in $E_R$ and $k_R$ (see Fig. S6).
However, one may expect that the buckling height in an experimentally grown sample 
should be rather close to the optimized value by our calculations.

In table~\ref{table1}, we summarize the Rashba parameter $\alpha_R$, energy offset $E_R$ and momentum offset $k_R$ for the systems studied.
$E_R$ and $k_R$ are for the bands marked in Figs.~\ref{fig2} and \ref{fig3}. 
We also show the SO splitting parameters of the heavy-element doped nonpolar surface Bi/Ag(111) and the surface of polar semiconductor BiTeI for comparison.
For the alloy systems SbBi/Al$_2$O$_3$(0001) and PbBi/Al$_2$O$_3$(0001),
$E_R$ is over 640 meV along -K-$\Gamma$-K, which is more than three times of that for Bi/Ag(111) and six times of that for the surface of BiTeI.
While $k_R$ is almost three times of that for Bi/Ag(111) and one order of magnitude larger than BiTeI.
Consequently, an unprecedented large Rashba parameter $\alpha_R$ is obtained for both SbBi/Al$_2$O$_3$(0001) and PbBi/Al$_2$O$_3$(0001).
Compared to the isolated system, e.g., SbBi, interfacing with Al$_2$O$_3$(0001) enhances $E_R$ by a factor of three 
and enhances $\alpha_R$ roughly by a factor of two. 
Our results thus demonstrate that the combination of ordered alloying and interfacing with a polar surface
can be an effective strategy to obtain a giant SO splitting in buckled monolayers.

    \begin{table*}
      \renewcommand{\arraystretch}{1.5}
      \begin{tabular}{lccccc}
    \hline
     Materials               & $E_R$  & $k_R$  & $\alpha_R$ & Identity & Reference  \\
     \hline 
\textbf{Heavy-element doped nonpolar surface}  \\
         Bi/Ag(111)          &  200   &  0.13  &   3.05     &  Mixed   &  \onlinecite{Bi-Ag111}\\
\textbf{Polar surface} \\
   BiTeI                     &  108   &  0.05  &   4.30     &  Ideal   &  \onlinecite{BiTeX}\\
\textbf{Homolayer on polar surface} \\
   Au/InSe(0001)             &   -    &    -   &   0.45     &  Ideal   &  \onlinecite{Au-InSe_2018}\\
Bi/Al$_2$O$_3$(0001)         &  160   &  0.20  &   1.61     &  Ideal   &  This work \\
Pb/Al$_2$O$_3$(0001)         &  266   &  0.13  &   4.14     &  Ideal   &  This work \\
\textbf{Isolated ordered alloy monolayer} \\
SbBi                         &  150   &  0.21  &   1.40     &  Ideal   &  This work \\
\textbf{Ordered alloy monolayer on polar surface} \\
SbBi/Al$_2$O$_3$(0001)       &  641   &  0.36  &   3.55     &  Ideal   &  This work \\
PbBi/Al$_2$O$_3$(0001)       &  741   &  0.34  &   4.38     &  Ideal   &  This work \\
    \hline 
      \end{tabular}
      \caption{SO splitting in buckled monolayers on Al$_2$O$_3$(0001) and selected materials from literature.
      $E_R$ and $k_R$ are the Rashba energy offset (meV) and momentum offset ($\AA^{-1}$) for the bands marked in Figs.~(\ref{fig2}) and (\ref{fig3}).
      $\alpha_R$ is the Rashba parameter (eV $\AA$) calculated by $\alpha_R=2E_R/k_R$.
       }
      \label{table1}
    \end{table*}
    \begin{table}
      \renewcommand{\arraystretch}{1.5}
      \begin{tabular}{ccccc}
    \hline
      Materials           & $\Delta H_{p_x,p_x}$ & $\Delta H_{p_y,p_y}$ & $\Delta H_{p_z,p_z}$  &  $\Delta z$ \\
     \hline 
   Pb/Al$_2$O$_3$(0001)   &  0.83                &  0.83                &    1.32               &   1.30      \\
   Bi/Al$_2$O$_3$(0001)   &  0.58                &  0.58                &    0.73               &   1.50      \\
 isolated SbBi            &  0.20                &  0.20                &    0.68               &   1.25      \\
 SbBi/Al$_2$O$_3$(0001)   &  1.07                &  1.07                &    1.31               &   1.25      \\
 isolated PbBi            & -0.46                & -0.46                &   -0.68               &   1.37      \\
 PbBi/Al$_2$O$_3$(0001)   &  0.37                &  0.37                &    0.64               &   1.37      \\ 
    \hline
      \end{tabular}
      \caption{Effects of alloying and interfacing on the on-site hopping parameters ($H_{\alpha,\alpha}$) 
      for the monolayers on Al$_2$O$_3$(0001). 
      $\Delta H_{\alpha,\alpha}$ (eV) is the difference in the on-site hopping terms for orbital $\alpha$
      between the two atoms in the overlayer,
      which takes the atom binding to the oxygen atoms as the reference. 
      $\Delta z$ (in the unit of \AA) denotes the buckling in the overlayer.}
      \label{table2}
    \end{table}

The large SO splittings are originated from the special geometry that induces a large perpendicular potential gradient through the overlayer. 
Within the tight-binding approximation, the effect of the perpendicular potential gradient goes into on-site Hamiltonian matrix elements.
We perform analyses on the Hamiltonian matrix elements as derived from the linear combination of atomic orbital (LCAO) calculation\cite{LCAO_openmx},
which reproduced the band structures shown in Figs.~\ref{fig2} and \ref{fig3} (see Fig. S7).
Table~\ref{table2} lists the differences in the on-site Hamiltonian matrix elements for the $p$-orbital between the two atoms in the overlayer, 
denoted as $\Delta H_{\alpha,\alpha}$, for which the atom binding to the oxygen atom is taken as the reference.
For the free-standing Bi-1L and Pb-1L, $\Delta H_{\alpha,\alpha}$ are zeros,
which become greater than 0.5 eV upon supported on Al$_2$O$_3$(0001).
The interface-induced perpendicular electric field through the supported Bi-1L is about 0.5 eV/\AA,
estimated by $\Delta H_{p_z,p_z}$/$\Delta z$. 
The estimated $E_z$ for Pb-1L is doubled of that for Bi-1L.
This trend is consistent with that $\alpha_R$ for Pb/Al$_2$O$_3$(0001) is much larger than that for Bi/Al$_2$O$_3$(0001) (see Table~\ref{table1}).
The difference between $\Delta H_{p_x,p_x}$ ($\Delta H_{p_y,p_y}$) and $\Delta H_{p_z,p_z}$ 
is attributed to the two-dimensional nature of the interface structure.

For the ordered alloy systems SbBi and PbBi,
the structure naturally gives a difference in the on-site Hamiltonian matrix elements,
which results in SO splittings in the electronic bands (Fig. S8). 
From Table~\ref{table2} one can see that $\Delta H_{\alpha,\alpha}$ are enhanced by over 0.6 eV for the supported SbBi.
For PbBi, interfacing reverses the potential gradient,
which leads to changes in $\Delta H_{\alpha,\alpha}$ over 0.8 eV for $p$ orbital.
Consequently, the SO splitting is significantly enhanced by interfacing  
compared to those for the isolated monolayer alloy.

In addition to the giant SO splittings, some of our systems may have nontrivial topological properties.
We have calculated the evolution of the Wannier function center based on the method proposed in Ref.\onlinecite{wannier_center}.
Our results reveal that Pb/Al$_2$O$_3$(0001) and Bi/Al$_2$O$_3$(0001) have an odd $Z_2$ number (Fig. S9),
whereas SbBi/Al$_2$O$_3$(0001) has an even $Z_2$ number (not shown).
We further performed calculations of edge states by making the surface monolayer into a nanoribbon.
Our calculations show that there are gapless edge states for Pb/Al$_2$O$_3$(0001) and Bi/Al$_2$O$_3$(0001) (Fig. S9).
Thus, these two systems are expected to be topologically nontrivial.

Lastly, we discuss the experimental feasibility of our systems.
The layered crystal structure of bismuth favors the growth of Bi-1L,
which has been obtained on several semiconducting substrates such as Bi$_2$Te$_3$ and Bi$_2$Se$_3$ by MBE growth 
\cite{Bi_Bi2Te3_2012,Bi_Bi2Se3_2017,Bi_Bi2Se3_2018}.
While the buckled honeycomb structure of Pb was predicted to be energetically lower than the planar one\cite{Pb_2016}.
Alloy systems such as Sb$_{1-x}$Bi$_x$ and Pb$_{1-x}$Bi$_x$ have been grown on Ag(111)
\cite{BiPb_2008,BiPb_2009,BiSb_BiPb_2011}.
Moreover, Pb-based and Bi-based ordered alloys on surfaces,
e.g., Tl$_3$Pb/Si(111) and Sn$_2$Bi/Si(111),
have also been obtained by recent experiments\cite{TlPb_2015,Sn2Bi_2018}.
Regarding the substrate, Al$_2$O$_3$(0001) has been extensively used as a substrate for the growth of thin films. 
For instance, it has been used for the growth of silicene\cite{Si_Al2O3_2018},
confirming our early DFT prediction\cite{Si_Al2O3_2016}.
Our calculations show that the binding energy ($E_b$) is about 0.25 eV/Bi for Bi/Al$_2$O$_3$(0001), 
much larger than that for Bi/Bi$_2$Te$_3$ (0.10 eV/Bi), a system already obtained in laboratory\cite{Bi_Bi2Te3_2012}.
While for Pb/Al$_2$O$_3$(0001) and the alloy systems, $E_b$ are larger than 0.50 eV/atom,
favoring the monolayer structure.
For isolated SbBi, a previous study found that it is both dynamically and thermally stable\cite{BiSb_2017}.
We further performed an \textit{ab initio} molecular dynamics simulation (T = 500K) for the supported system, i.e., SbBi/Al$_2$O$_3$(0001),
which shows that the structure is also thermally stable (Fig. S10). 
Therefore, the growth of our systems can be highly feasible.

\section*{CONCLUSION}
In summary, we have proposed a strategy that combines the surface alloying and interface engineering to manipulate the SO splitting in 2D materials. 
We have illustrated the idea in low-buckled hexagonal monolayers, e.g., Bi-1L, Pb-1L and their alloys SbBi and PbBi,
supported on Al$_2$O$_3$(0001) by DFT calculations.
Our calculations show giant Rashba-like SO splittings in these interface structures.
In particular, the Rashba energies and momentum offsets of the split states 
for SbBi/Al$_2$O$_3$(0001) and PbBi/Al$_2$O$_3$(0001) are roughly three times of those for Bi/Ag(111).
Our study thus provides an effective way of manipulating the SO splitting in layered 2D materials 
 for potential applications in spintronics and the study of Majorana Fermions in solids. 

\section*{METHODS}
Our calculations were performed using the Vienna Ab Initio Simulation Package\cite{kresse1996}. 
The interface structure is modeled in terms of a repeated slab, 
separated from its periodic images by 10 \AA vacuum regions. 
We note that the lattice mismatch strain can be effectively relieved by adjusting the buckling height of the overlayer
\cite{Bi_Bi2Se3_2017,Bi_Bi2Se3_2018}.
Therefore, the interface structure in our modeling contains only one unit cell for both the overlayer and the substrate.
The pseudopotentials were constructed by the projector augmented wave method\cite{kresse1999}. 
Van der Waals dispersion forces between the adsorbate and the
substrate were accounted for through the optPBE-vdW functional by using the vdW-DF method\cite{klimes2010}. 
A 11$\times$11 $\Gamma$-centered Monkhorst-Pack $k$-point mesh was used to sample the surface Brillouin zone. 
A plane-wave energy cutoff of 400 eV was used for all the calculations. 
The overlayer atoms and the surface Al and O atoms are fully relaxed until the residual forces are less than 0.001 eV/\AA.

\section*{FUNDING}
M.C. thanks financial support by the National Natural Science Foundation of China (Grant Nos. 11774084 and 91833302)
and the Project of Educational Commission of Hunan Province of China (Grant No. 18A003).
F.L. is supported by U.S. Department of Energy-Basic Energy Sciences (Grant No. DE-FG02-04ER46148).

\section*{AUTHOR CONTRIBUTIONS}
M.C. proposed the project and carried out the calculations. 
All authors contributed to the analysis of the data and the manuscript writing.

\section*{Conflict of interest statement}
None declared.

\bibliography{references}
\bibliographystyle{apsrev4-1}
\end{document}